\documentclass[useAMS,usenatbib]{mn2e}
\usepackage{graphics,epsfig,amsmath,amssymb,amstext,txfonts,array,color}
\usepackage{graphicx}
\usepackage[T1]{fontenc}

\usepackage{color}

\newcommand\simlt{\lower.5ex\hbox{$\; \buildrel < \over \sim \;$}}
\newcommand\simgt{\lower.5ex\hbox{$\; \buildrel > \over \sim \;$}}

\title[Reconfinement of highly magnetized jets]
{Reconfinement of highly magnetized jets: Implications for HST-1 in M87}

\author[Amir Levinson, Noemie Globus]
  {A.~Levinson,$^1$\thanks{E-mail: levinson@wise.tau.ac.il}
  N.~Globus,$^2$\\
  $^1$Raymond and Beverly Sackler School of Physics \& Astronomy, Tel Aviv University, Tel Aviv 69978, Israel\\
  $^2$ Racah Institute of Physics, The Hebrew University of Jerusalem, 91904 Jerusalem, Israel}
\date{Released \today}

\pagerange{\pageref{firstpage}--\pageref{lastpage}} \pubyear{2002}

\def\LaTeX{L\kern-.36em\raise.3ex\hbox{a}\kern-.15em
    T\kern-.1667em\lower.7ex\hbox{E}\kern-.125emX}

\begin{document}
\label{firstpage}
\maketitle

\begin{abstract}
Stationary features are occasionally observed in AGN jets.  A notable example is the HST-1 knot in M87.
Such features are commonly interpreted as re-confinement shocks in hydrodynamic jets or
focusing nozzles in Poynting jets.  In this paper we compute the structure and Lorentz factor of a highly magnetized jet 
confined by external pressure having a profile that flattens abruptly at some radius.
We find the development of strong oscillations upon transition from the steeper to the flatter pressure 
profile medium.   Analytic formula is derived for the location of the nodes of these oscillations.  We apply the 
model to the M87  jet and show that if the jet remains magnetically dominated up to sub-kpc scales, then focusing is 
expected.  The location of the HST-1 knot can be reconciled with recent measurements of the 
pressure profile around the Bondi radius if the jet luminosity satisfies $L_j\simeq10^{43}$ erg/s.   However, 
we find that magnetic domination at the collimation break implies a Lorentz factor in excess of $10^2$, atypical
to FRI sources.    A much lower value of the asymptotic Lorentz factor would require substantial loading
close to the black hole. In that case HST-1 may be associated with a collimation nozzle of a hydrodynamic flow.
\end{abstract} 

\section{Introduction}

HST-1 is a stationary radio feature associated with the sub-kpc scale jet in M87 (Biretta et al. 1999). The knot 
is located at a projected distance of 60 pc (0.86$\arcsec$) from the central engine, and is 
known to be a region of violent activity.  Subfeatures moving away from the main knot of 
the HST-1 complex at superluminal speeds have been detected (Biretta et al. 1999; Cheung et al. 2007, Giroletti et al. 2012), 
indicating that a highly relativistic flow is passing through this region. Moreover, the reported 
variability of the resolved X-ray and optical emission from HST-1 implies that the (beaming corrected) size of the emission region is much 
smaller than the distance between the HST-1 knot and the central black hole (e.g., Cheung et al. 2007), 
suggesting a strong focusing of the flow at this location.   Stationary radio features, similar to 
HST-1, are quite common in radio loud  AGNs (e.g., Coehn, et al. 2014) and may have the same origin.

It has been proposed that the stationary knot in the HST-1 complex is associated with a recollimation 
shock that forms either, due to a  change in the gradient of the external pressure
(Stawarz et al. 2006; Bromberg \& Levinson 2009, hereafter ST06 and BL09), or by hoop stresses of a subrelativistic, magnetized 
wind surrounding the spine, as in the two-component MHD model of Garcia et al. (2009; see also Nakamura et al. 2010).
The recollimation shock scenario is further supported by the sudden change in the collimation profile observed 
at the location of HST-1 (Nakamura \& Asada 2013).
In this picture, the rapid variability of the HST-1 emission and the ejection of superluminal subknots from the HST-1 
complex are associated with internal shocks produced by self-reflection of the converging flow  at the axis, 
just downstream of the recollimation nozzle (Bogovalov \& Tsinganos 2005; 
Levinson \& Bromberg 2008; Nakamura et al. 2010, 2014).

\begin{figure*}
\includegraphics[width=14.5cm]{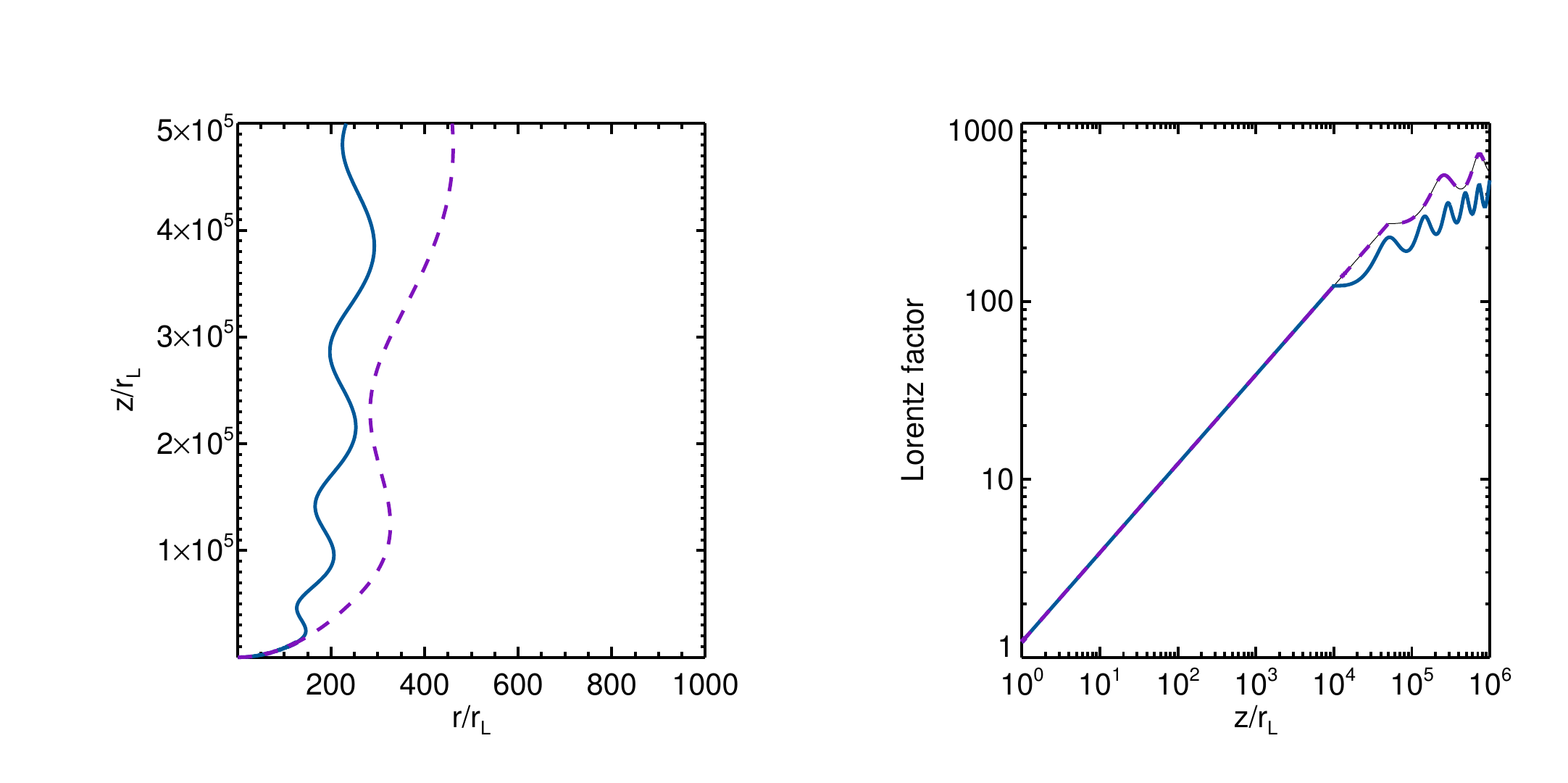}
\caption{\label{f1} Jet profile (lef panel) and Lorentz factor (right panel) computed using 
a confining pressure of the form ${\cal P}_{ext} = A\, z^{-2}$ at $z<z_{tr}$ 
and ${\cal P}_{ext} = (A/z_{tr})\, z^{-1}$ at $z>z_{tr}$, for $z_{tr}=10^4$ (solid line) and $z_{tr}=5\times10^4$ (dashed line).}
\end{figure*}

A question of interest is whether the M87 jet is highly or weakly magnetized at the recollimation nozzle.   
ST06 considered a hydrodynamic jet, and proposed that HST-1 reflects the location at which the 
reconfinement shock reaches the jet axis.  They did not attempt to reproduce the entire collimation profile, 
but merely to demonstrate that convergence of the shock occurs under the conditions inferred from observations. 
BL09 have shown that strong focusing of a hydrodynamic jet at the reflection point is 
anticipated if the plasma behind the oblique shock cools rapidly.   Both, ST06 and BL09 attribute the location of 
the nozzle to a change in the external pressure profile.  In the two-component MHD model 
(Bogovalov \& Tsinganos 2005, Garcia et al. 2009), on the other hand, the focusing of the jet is not related to 
the properties of the ambient medium, but rather to the dynamics of a magnetized spine-sheath
structure.  This model can reproduce the entire collimation profile upon appropriate choice of model 
parameters.   However, it requires the existence of an extended  MHD disk wind with a regular magnetic 
field which, in our view, is highly questionable.  The observed collimation profile within the Bondi radius 
can also be reproduced if the inner, Poynting jet, is confined by a hydrodynamic (weakly magnetized) disk 
wind (Globus \& Levinson 2016, hereafter GL16).   Whether it can also account for the 
change in structure across the Bondi radius is one of the issues addressed in this paper.   

A highly magnetized jet is expected to undergo strong oscillations when encountering a medium
with a flat pressure profile, $p_{ext}(z)\propto z^{-\kappa}$, $\kappa<2$ (Lyubarsky 2009, Komissarov et al. 2015, Mizuno et al. 2015).
These oscillations occur in cases where the jet is not in its equilibrium state when encountering the 
confining medium (Lyubarsky 2009), and can be ascribed to a standing magnetosonic wave.  In reality,
substantial focusing of the jet should lead to a rapid growth of current-driven
internal kink modes, and the subsequent dissipation of the magnetic field (Bromberg \& Tchekhovskoy 2016, Singh et al. 2016).  
Thus, focusing of a Poynting jet can, in principle, produce stationary jet features with observational
imprints similar to those observed in M87 and other AGNs.

In this paper we address the following question: can HST-1 in M87 be produced by the focusing of 
a Poynting-flux dominated jet in a way that satifies all observational constraints? And if so, what are the 
implications for jet dynamics?

\section{Model description and results}\label{sec:model}
In the following, we use cylindrical coordinates ($r$,$\phi$,$z$), 
with the $z$ axis aligned with the jet symmetry axis.
We seek to compute the profile and Lorentz factor of a strongly magnetized jet confined by external 
pressure, $p_{ext}(z)$, with a changing profile.
To be concrete, we suppose that the gradient of the external pressure changes abruptly across a transition 
point, henceforth denoted by $z_{tr}$, from a steeper profile, $p_{ext}(z)=p_<(z)$ at $z<z_{tr}$, to a flatter profile, $p_{ext}(z)=p_>(z)$ 
at $z>z_{tr}$.   
In what follows, we normalize  the external pressure to the fiducial value
\begin{equation}
p_0=\frac{2L_j}{3\pi c\,r_L^2}=200\, L_{j44}\,(r_L/r_s)^{-2}\, \left(\frac{M_{BH}}{6\times10^9\,M_\odot}\right)^{-2}\quad {\rm dyn\, cm^{-2}},\label{p_0_norm}
\end{equation}
where $L_j=10^{44}L_{j44}$ erg/s is the total jet power, $r_L=c/\Omega$ is the radius of the light cylinder,
and $r_s=1.8\times10^{15}\, (M_{BH}/6\times10^{9} M_\odot)$ cm is the Schwarzschild radius, and denote ${\cal P}_{ext}(z)=p_{ext}(z)/p_0$.
The shape of the jet radius, $r_j(z)$, is computed by employing the asymptotic transfield equation 
derived in Lyubarsky (2009), which
holds well above the light cylinder in regions where the jet is confined.   
Upon substituting  $L_j=(B_0^2/4\pi)c\pi r_L^2$,
where $B_0$ is the strength of the magnetic field at $r_L$, and using Equation (\ref{p_0_norm}), 
it reads:
\begin{equation}
\frac{d^2 r_j(z)}{dz^2}=\frac{1}{[r_j(z)]^3}- {\cal P}_{ext}(z) r_j(z).
\label{asymptotic_jet}
\end{equation}
Henceforth all lengths are measured in units of $r_L$ unless otherwise specified.
The corresponding Lorentz factor at the jet boundary is given by (Lyubarsky 2009)
\begin{equation}
\Gamma(z)=3^{1/4}{\cal P}_{ext}^{-1/2}r_j^{-1}.
\label{Gamma_j}
\end{equation}
It is worth noting that in general the Lorentz factor is not uniform across the jet.   It increases from its minimum 
value on the jet axis to its maximum value near the boundary (Lyubarsky 2009).

Equation (\ref{asymptotic_jet}) is integrated numerically using the prescribed pressure ${\cal P}_{ext}(z)$. 
For illustration we consider a pressure profile of the form ${\cal P}_{ext} = A\, z^{-2}$ at $z<z_{tr}$ 
and ${\cal P}_{ext} = B\, z^{-\kappa}$, $\kappa<2$, at $z>z_{tr}$, with $B=A\, z_{tr}^{\kappa-2}$.  
As discussed in Lyubarsky (2009), for the former profile solutions with $A<1/4$ correspond to non-equilibrium collimation, 
whereas $A>1/4$ corresponds to an intermediate case.   For the intermediate case ($A > 1/4$) we choose the initial values of 
our integration, at $z=1$, to reproduce the non-oscillatory analytic solution derived in Lyubarsky (2009),
$r_j=z^{1/2}/(A-1/4)^{1/4}$, in the region $z<z_{tr}$.  
In the non-equilibrium regime ($A<1/4$), the solution in the region $z<z_{tr}$ always approaches 
$r_j\propto z^{\eta}$, $\eta=(1+\sqrt{1-4A})/2$, at sufficiently large $z$ 
(Komissarov et al. 2009; Lyubarsky 2009).   In this regime our initial conditions are simply $r_j(z=1)=r_{j0}$, $dr_j/dz|_{z=1}=\eta\, r_{j0}$.
We then integrate Equation (\ref{asymptotic_jet}) from $z=1$ across the transition point  to $z>>z_{tr}$, using the appropriate 
pressure profile in each region.

\begin{figure*}
\includegraphics[width=14.5cm]{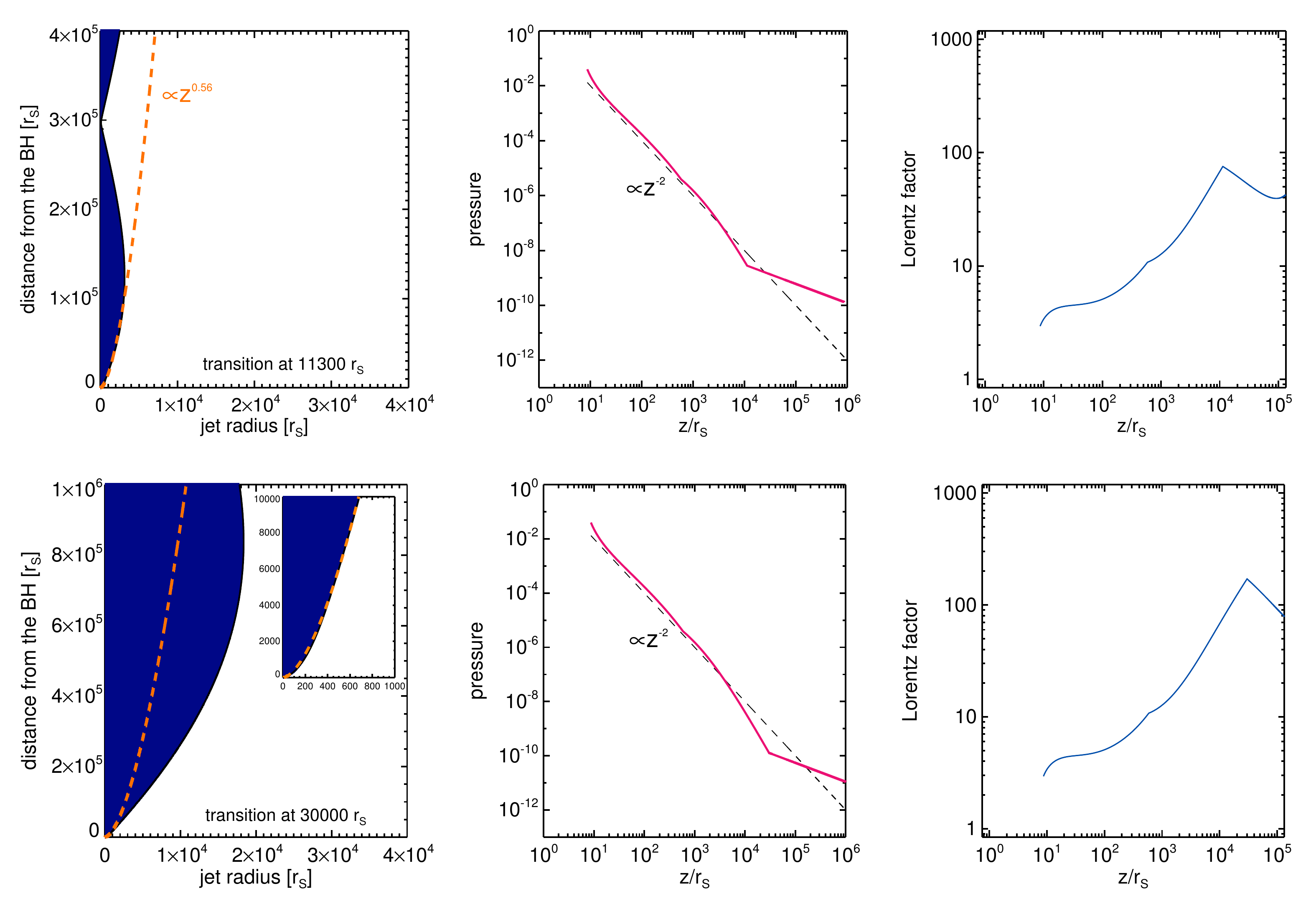}
\caption{\label{f2} Profiles of jet radius (left panels), pressure (middle panels) and Lorentz factor (right panels) for the M87 jet model. 
The upper panels correspond to a transition radius of $z_{tr}=1.1\times10^4\, r_s$ and the bottom panels to $z_{tr}=3\times10^4\, r_s$. 
In the left panels the vertical axis marks the distance to the black hole, and the horizontal axis the jet radius.  The pressure 
is normalized to the fiducial value given in Equation (\ref{p_0_norm}).  The dashed red line in the left panels marks 
the profile $r_j\propto z^{0.58}$ observed below the collimation break (Nakamura \& Asada 2013). The inset in the lower left panel is a zoom-in 
of the inner jet region, presented here for clarity.} 
\end{figure*}

An example is shown in Figure \ref{f1} for $\kappa=1$ and two different values of the transition point;
$z_{tr}=10^4$ (solid line) and $z_{tr}=5\times10^4$ (dashed line). 
As seen, the jet undergoes strong oscillations upon encountering the 
flat pressure profile medium at $z>z_{tr}$.  The amplitude of these oscillations increases with 
decreasing pressure at the transition point (see appendix).  The wavelength of the oscillations
depend on the pressure at the transition point, ${\cal P}_{ext}(z_{tr})$, and the slope $\kappa$.
The derivation in the appendix yields an estimate for the location of the nodes: 
\begin{equation}
z_n=\left[\frac{(n+1/2+\alpha-n\kappa/2-\alpha\kappa/2-\kappa/8)\pi}{z_{tr}\sqrt{{\cal P}_{ext}(z_{tr})}}
\right]^{1/(1-\kappa/2)}\, z_{tr},\quad n=0,1,2,...,
\label{nodes}
\end{equation}
restricted to $z_n>z_{tr}$, where $\alpha$ is a real number between $0.5$ and $1$, that depends 
on the phase of the solution at $z=z_{tr}$.  At low pressures, viz., $z_{tr}\sqrt{{\cal P}_{ext}(z_{tr})}<<1$,
$\alpha$ approaches either $0.5$ or $1$ (see appendix).
Equation (\ref{nodes}) is in excellent agreement with our numerical results.  
As noted above, in reality rapid dissipation of the magnetic 
field is expected at the first focusing point, that may alter the subsequent evolution of the jet.
By employing Equations (\ref{p_0_norm}) and (\ref{nodes}) 
and the relation ${\cal P}_{ext}(z_{tr})(z_{tr}/z_0)^{\kappa}={\cal P}_{ext}(z_0)$, and adopting $\alpha=1/2$, 
we obtain the pressure at the first focusing point ($n=0$):
\begin{equation}
p_{ext}(z_{0})= 7\times 10^{-8}(1-3\kappa/8)^2L_{j44}\left(\frac{z_{0}}{100\, \rm pc}\right)^{-2}\quad {\rm dyn\, cm^{-2}},
\end{equation}
independent of the black hole mass.

\subsection{Application to M87}

We now apply the model outlined above to the HST-1 knot in M87.    This stationary feature, located at a projected distance of
about 60 pc from the core, appears to be associated with a change in the collimation profile of the jet (Nakamura \& Asada 2013), as anticipated 
in a collimation nozzle.   The apparent speed of superluminal motions on kpc scales constraint the 
viewing angle to the M87 jet to be less than about $20^\circ$.  Below we adopt a viewing angle of $18^\circ$, that yields an
actual distance of 200 pc for HST-1, or $3\times10^5\, r_s$  for a black hole mass of $M_{BH}=6\times10^9\, M_\odot$.
The collimation profile of the jet has been probed down to about $10\, r_s$.  It maintains a parabolic shape 
($r_j\propto z^{0.58}$) all the way up to the transition point, above which it becomes roughly conical.    This readily implies 
that the profile of the confining pressure  in this region is roughly $p_{ext}\propto z^{-2}$.   
As shown elsewhere (GL16),  the collimation profile of the jet below the transition point can be reproduced through a
collision of a disk wind with the jet, for a certain choice of  jet-to-wind power ratio.   In this model the confining pressure
in Equation (\ref{asymptotic_jet}) is obtained self-consistently from the compression of the disk wind near the jet boundary.   
Although the origin of the confining pressure is not important for the present analysis, as a matter of convenience 
we use the method outlined in  GL16  to integrate Equation (\ref{asymptotic_jet}) below the transition point.   

Above the transition point we invoke an external pressure of the form $p_>(z)=p_{tr}(z/z_{tr})^{-\kappa}$,
with $\kappa \simeq 0.7$ and $p_{tr}$ determined by the requirement that the pressure is continuous across the transition. 
This profile is inferred  from X-ray observations on scales $0.2 - 10 $ kpc (Russell et al. 2015), 
with a normalization $p(z=100 pc)=10^{-9}$ dyn cm$^{-2}$.   We suppose that this profile extends down to $z_{tr}$. 
In the following calculations, we adopt the value $r_L=2.25\, r_s$ obtained in Globus \& Levinson (2014) for the outer 
light surface of an equatorial, force-free outflow from a Kerr black hole with a spin parameter $a=0.9$, 
assuming a rigid rotation of magnetic field lines with angular velocity $\Omega=\Omega_H/2$,
where $\Omega_H$ is the angular velocity of the black hole. 

Figure \ref{f2} shows the resultant jet radius, pressure profile and Lorentz factor, for $z_{tr}=1.1\times10^4\, r_s$ and $z_{tr}=3\times 10^4\, r_s$.
As can be seen, the location of the first focusing point is in good agreement with Equation (\ref{nodes}) upon substituting $\alpha=0.5$ and the
value of ${\cal P}(z_{tr})$ obtained from our integration (middle panels).   Note that $z$ and $r_j$ in Figure \ref{f2} are measured in units of $r_s$,
not $r_L$.   The value of $z_{tr}$ in the solution presented in the upper panels was chosen such that it reproduces the observed structure.  The
solution in the lower panels is exhibited for comparison.    Both solutions are essentially the same up to the transition  radius roughly.  
We find, quite generally, that the location of the focusing point can be reconciled with  the observed collimation break 
of the M87 jet provided $z_{tr}\simeq10$ pc  and ${\cal P}(z_{tr})\simeq1.8\times10^{-9}$ (see upper left panel), 
or using Equation (\ref{p_0_norm}) $p_{ext}(z_{tr})=7\times10^{-8}\, L_{j44}$ dyn cm$^{-2}$, independent of the black hole mass.   
This value is consistent with the ambient pressure inferred from X-ray observations (Russell et al. 2015) provided
\begin{equation}
L_{j44}=\left[\frac{p_{obs}(z=100\, \rm pc)}{7\times10^{-8}\, {\rm dyn\, cm^{-2}}}\right]\left(\frac{z_{tr}}{100\, \rm pc}\right)^{-0.7}
\simeq 0.07.
\end{equation}

The Lorentz factor exhibited in the right panels of figure \ref{f2} was computed for a magnetically dominated jet.   As seen, 
the observed collimation profile of the M87 jet implies that a Poynting dominated jet should accelerate to a Lorentz in excess 
of $10^2$ at  the focusing nozzle.    This value is much higher than the average value inferred from statistics of 
superluminal motions  (e.g., Kellermann et al. 2007).  Either the M87 jet is unique, or loading close to the black hole limits 
the  terminal Lorentz factor. For example,
$\Gamma_j=10$ is achieved at a distance of about $200\, r_s$ from the putative black hole. We note that the collimation profile 
is expected to be insensitive to the local magnetization parameter, but the properties of the collimation nozzle may depend on it. 
If saturation occurs well below the collimation break, then HST-1 may reflect an MHD (Nakamura et al. 2010) or a 
hydrodynamic (ST06, BL09) shock.

Recent kinematic measurements on scales $100 - 5000\, r_s$ reveal substantial stratification of the flow, with 
a range of Lorentz factors $\Gamma \sim 1 - 3$ (Asada et al. 2014; Mertens et al. 2016) .   These values are somewhat 
lower than the Lorentz factor exhibited in the upper right panel of figure \ref{f2}.   However, one should keep in mind that  
the profile shown in figure \ref{f2} corresponds to the Lorentz factor at the jet boundary.   In general 
the Lorentz factor vary across the jet, and has lower values away from the boundary (see comment below 
Equation \ref{Gamma_j}), so one must be cautious in interpreting observational results.   Moreover, it could be that
the emission observed on those scales originates from a slower sheath flow (e.g., Moscibrodzka et al. 2016; GL16), as
seems to be indicated by the detection of subluminal motions in this region.

\section{Summary and conclusions}

A Poynting-flux dominated jet undergoes strong focusing upon encountering a confining medium with 
a flat pressure profile, $p_{ext}\propto z^{-\kappa}$, $\kappa<2$.  At the focusing nozzle, rapid 
growth of internal kink modes, followed by dissipation of the magnetic field is anticipated (Bromberg \& Tchekhovskoy 2016, Singh et al. 2016). 
This can lead to the appearance of stationary emission features, as occasionally seen in radio loud AGNs.

In case of M87 we find that the entire collimation profile can be reproduced 
if the confining pressure profile changes from approximately $p_{ext}\propto z^{-2}$ to the profile observed on 
scales $0.1$-$10$ kpc, $p_{ext}\propto z^{-0.7}$.    The location of the HST-1 knot can be reconciled with 
the measured pressure provided that the jet power on those scales is $L_j\simeq 10^{43}$ erg/s.   
This value is significantly lower than the power of the large
scale jet  estimated from various observations (Bicknell \& Begelman 1996; de Gasperin et al. 2012), 
but is consistent with values obtained in recent GRMHD 
simulations (Moscibrodzka et al. 2016).   As pointed out in the later reference, the jet power may be intermittent 
on timescales much shorter than the age of the system,
roughly 30 Myr, and so it is conceivable that it is currently in a low state,  well below the average power estimated on large scales.
It is also found that if the jet remains magnetically dominated up to the focusing nozzle, here associated with the stationary feature 
in the HST-1 complex, then its Lorentz factor should be in excess of several hundreds.    Such a high Lorentz factor is atypical 
to FRI sources, as inferred from statistics of superluminal motions (e.g., Kellermann et al. 2007). 
A much lower value of the asymptotic Lorentz factor would require substantial loading
close to the black hole.  We do not expect significant alteration of the collimation profile  of the loaded jet, 
but the nature of HST-1 may be different.

The relative proximity of HST-1 to the black hole implies that the confining pressure around the Bondi radius, 
that produces the collimation break, is unlikely to be supported
by the cocoon produced by the relativistic jet, as hinted by Tchekhovskoy \& Bromberg (2016).  
Whether the interaction of disk winds with the ambient medium can account for the observed pressure profile on those scales 
remains to be investigated.  If the jet is indeed collimated by disk winds, as proposed recently (GL16), 
then one naively anticipates strong shocks to form by virtue of the interaction of the disk wind with the flat density galaxy core. \\

We thank Masanori Nakamura for enlightening discussions and useful comments.
AL acknowledges the support of The Israel Science Foundation (grant 1277/13).
NG acknowledges the support of the I-CORE Program of the Planning and Budgeting Committee and The Israel Science Foundation (grant 1829/12) 
and the Israel Space Agency (grant 3-10417).

\appendix
\section{Analysis of jet oscillations}

For the external pressure invoked in the region $z>z_{tr}$ in section \ref{sec:model}, 
${\cal P}_{ext}(z)=Bz^{-\kappa}$, $\kappa<2$,
the transfield equation admits the analytic solution (Lyubarsky 2009) 
\begin{equation}
r_j=\sqrt{\frac{2-\kappa}{\pi}}\left(\frac{z^\kappa}{B}\right)^{1/4}\Phi(z)
\end{equation}
here 
\begin{equation}
\Phi(z)=\frac{1}{C_1}\cos^2S+C_1\left(C_2\cos S+\frac{\pi}{2-\kappa}\sin S\right)^2,
\end{equation}
and 
\begin{equation}
S=\frac{2\sqrt{B}}{2-\kappa}z^{1-\kappa/2}-\frac{(4-\kappa)}{(2-\kappa)}\frac{\pi}{4}.
\end{equation}
This solution needs to be matched to the solution invoked in section \ref{sec:model} in the region $z<z_{tr}$.  
The coefficients $C_1$ and $C_2$ are then determined from the requirement that  $r_j(z)$ and its derivative are continuous 
at the transition point $z_{tr}$.   For $z_{tr}>>1$ the minima of $r_j(z)$ coincide to a good approximation with the minima of
the function $\Phi(z)$.  The minima of $\Phi$ occur at the phases $\varphi_n=(\alpha+n)\pi$, $n=0,1,2,...$, 
with $1/2<\alpha<1$, given by
\begin{equation}
\cos \varphi_n= \frac{(-1)^{n+1}}{1+(b+\sqrt{b^2+1})^2}
\end{equation}
in terms of
\begin{equation}
b=\frac{2-\kappa}{2\pi C_1^2C_2}+\frac{(2-\kappa)C_2}{2\pi}-\frac{\pi}{2(2-\kappa)C_2}.
\end{equation}
The location of the nodes is determined from the relation $S(z_n)=\varphi_n$.  In the limit $|b|>>1$ one finds $\cos\varphi_n\rightarrow 0$ for 
$b>0$, corresponding to $\alpha=1/2$, and $\cos\varphi_n\rightarrow (-1)^{n+1}$ for $b<0$, corresponding to $\alpha=1$.


\begin{thebibliography}{99}
\bibitem[Asada et al. (2014)]{2014ApJ...781L...2A} Asada K., Nakamura M., Doi A., Nagai H., Inoue M., 2014, ApJ, 781, L2 
\bibitem[Bicknell \& Begelman (1996)]{BB96}  Bicknell, G. V., \& Begelman, M. C. 1996, ApJ, 467, 597
\bibitem[Biretta et al. (1999)]{BSM99} Biretta, J. A., Sparks, W. B., \& Macchetto, F. 1999, ApJ, 520,621
\bibitem[Bogovalov \& Tsinganos (2005)]{BT05}  Bogovalov, S., \& Tsinganos, K. 2005, MNRAS, 357, 918
\bibitem[Bromberg \& Levinson (2009)]{BL09} Bromberg, O. \& Levinson, A. 2009, ApJ, 699, 1274
\bibitem[Bromberg \& Tchekhovskoy (2016)]{BT16}  Bromberg O. \& Tchekhovskoy, A. 2016, MNRAS, 456, 1739
\bibitem[Cheung et al. (2007)]{Chetal07}  Cheung, C. C., Harris, D. E., \& Stawarz, L. 2007, ApJ, 663, L65
\bibitem[Cohen et al. (2014)]{Coetal14} Cohen M. H., Meier, D. L., Arshakian T. G., Homan, D. C.,  Hovatta, T.,  Kovalev, Y. Y.,
Lister, M. L., Pushkarev, A. B., Richards, J. L., Savolainen T. 2014, ApJ, 787, 151
\bibitem[de Gasperin et al. (2012)]{2012A&A...547A..56D} de Gasperin F., et al., 2012, A\&A, 547, A56 
\bibitem[Garcia et al. (2009)]{Garelat09}  Gracia, J.; Vlahakis, N.; Agudo, I.; Tsinganos, K.; Bogovalov, S. V. 2009, ApJ, 695, 503
\bibitem[Giroletti et al. (2012)]{Grelat12}  Giroletti, M. et al. 2012, A\&A, 538, 10
\bibitem[Globus \& Levinson (2016)]{GL16} Globus, N.; Levinson, A. 2016, MNRAS, 461, 2605
\bibitem[Kellermann et al. (2007)]{Kletal07} Kellermann, K. I. etal. 2007, A\&SS, 311, 231
\bibitem[Komissarov et al. (2009)]{Komissarov09} Komissarov S., Vlahakis N., Konigl A., Barkov M. V., 2009, MNRAS, 394, 1182
\bibitem[Komissarov et al. (2015)]{Komissarov15}  Komissarov, S. S.; Porth, O.; Lyutikov, M.  2015,  ComAC, 2, 9
\bibitem[Levinson \& Bromberg (2008)]{LB08} Levinson A. \& Bromberg, O. 2008, IJMPD, 17, 1603
\bibitem[Lyubarsky (2009)]{Lyu09} Lyubarsky, Y. 2009, ApJ, 698, 1570
\bibitem[Mertens et al. (2016)]{2016arXiv160805063M} Mertens F., Lobanov A.~P., Walker R.~C., Hardee P.~E., 2016, arXiv, arXiv:1608.05063 
\bibitem[Mizuno et al. (2015)]{Mietal15} Mizuno, Y., Gomez, J. L., Nishikawa, K.-I., Meli, A., Hardee, P. E., and Rezzolla, L. 2015, ApJ, 809, 28
\bibitem[Mo{\'s}cibrodzka, Falcke, \& Shiokawa (2016)]{2016A&A...586A..38M} Mo{\'s}cibrodzka M., Falcke H., Shiokawa H., 2016, A\&A, 586, A38 
\bibitem[Nakamura, et al. (2010)]{Nketal10} Nakamura, M., Garofalo, D. \& Meier, D. L,  2010, ApJ, 721, 1783
\bibitem[Nakamura \& Asada (2013)]{2013ApJ...775..118N} Nakamura M., Asada K., 2013, ApJ, 775, 118 
\bibitem[Nakamura \& Meier  (2014)]{NM14} Nakamura, M. \& Meier, D. L.  2014, ApJ, 785, 152
\bibitem[Russell et al. (2015)]{Russell15} Russell, H. R., Fabian, A. C., McNamara, B. R., Broderick, A. E. 2015, MNRAS, 451, 588
\bibitem[Singh et al. (2016)]{Sietal16}  Singh, Chandra B.; Mizuno, Yosuke; de Gouveia Dal Pino, Elisabete M. 2016, ApJ, 824, 48
\bibitem[Stawarz et al. (2006)]{SAK06} Stawarz, L, Aharonian, F., Kataoka, J., et al. 2006, MNRAS, 370, 981
\bibitem[Tchekhovskoy \& Bromberg (2016)]{2016MNRAS.461L..46T} Tchekhovskoy A., 
Bromberg O., 2016, MNRAS, 461, L46 


\end{thebibliography}
\end{document}